\documentclass[sigconf,screen]{acmart}

\AtBeginDocument{%
  \providecommand\BibTeX{{%
    \normalfont B\kern-0.5em{\scshape i\kern-0.25em b}\kern-0.8em\TeX}}}

\setcopyright{acmcopyright}
\acmPrice{15.00}
\acmDOI{10.1145/3540250.3549152}
\acmYear{2022}
\copyrightyear{2022}
\acmSubmissionID{fse22main-p849-p}
\acmISBN{978-1-4503-9413-0/22/11}
\acmConference[ESEC/FSE '22]{Proceedings of the 30th ACM Joint European Software Engineering Conference and Symposium on the Foundations of Software Engineering}{November 14--18, 2022}{Singapore, Singapore}
\acmBooktitle{Proceedings of the 30th ACM Joint European Software Engineering Conference and Symposium on the Foundations of Software Engineering (ESEC/FSE '22), November 14--18, 2022, Singapore, Singapore}

\usepackage[linesnumbered,ruled,vlined]{algorithm2e}
\usepackage{multirow}
\usepackage{hhline}
\usepackage{colortbl}
\usepackage[caption=false]{subfig}
\usepackage{balance}

\newlength\mylen

\PassOptionsToPackage{table, dvipsnames}{xcolor} 

\newcommand{\tool}{\textsc{STRaP}}

\newcommand{\ccomment}[1]{\textcolor[rgb]{0.25,0.49,0.41}{#1}}
\newcommand{\ttt}[1]{\tt\small{#1}}

\begin{document}

\title{Scenario-Based Test Reduction and Prioritization for Multi-Module Autonomous Driving Systems}


\author{Yao Deng}
\affiliation{%
  \institution{Macquarie University}
  \city{Sydney}
  \state{NSW}
  \country{Australia}}
\email{yao.deng@hdr.mq.edu.au}

\author{Xi Zheng}
\authornote{Corresponding authors: Xi Zheng, Mengshi Zhang.}
\affiliation{%
  \institution{Macquarie University}
  \city{Sydney}
  \state{NSW}
  \country{Australia}}
\email{james.zheng@mq.edu.au}

\author{Mengshi Zhang}
\authornotemark[1]
\affiliation{%
  \institution{Meta}
  \city{Menlo Park}
  \state{CA}
  \country{USA}}
\email{mengshizhang@fb.com}

\author{Guannan Lou}
\affiliation{%
  \institution{Macquarie University}
  \city{Sydney}
  \state{NSW}
  \country{Australia}}
\email{guannan.lou@mq.edu.au}

\author{Tianyi Zhang}
\affiliation{%
  \institution{Purdue University}
  \city{West Lafayette}
  \state{IN}
  \country{USA}}
\email{tianyi@purdue.edu}


\begin{abstract}
When developing autonomous driving systems (ADS), developers often need to replay previously collected driving recordings to check the correctness of newly introduced changes to the system. However, simply  replaying the entire recording is not necessary given the high redundancy of driving scenes in a recording (e.g., keeping the same lane for 10 minutes on a highway). 
In this paper, we propose a novel test reduction and prioritization approach for multi-module ADS. First, our approach automatically encodes frames in a driving recording to feature vectors based on a driving scene schema. Then, the given recording is sliced into segments based on the similarity of consecutive vectors. Lengthy segments are truncated to reduce the length of a recording and redundant segments with the same vector are removed. The remaining segments are prioritized based on both the coverage and the rarity of driving scenes. We implemented this approach on an industry-level, multi-module ADS called Apollo and evaluated it on three road maps in various regression settings. 
The results show that our approach significantly reduced the original recordings by over 34\% while keeping comparable test effectiveness, identifying almost all injected faults. Furthermore, our test prioritization method achieves about 22\% to 39\% and 41\% to 53\% improvements over three baselines in terms of both the average percentage of faults detected (APFD) and TOP-K.    
\end{abstract}

\ccsdesc[500]{Software and its engineering~Software testing and debugging}
\keywords{Autonomous Driving, Testing Reduction, Test Prioritization, Regression Testing}

\maketitle

\section{Introduction}

Autonomous driving has been through quick development in recent years and found several new business models such as driverless taxis~\cite{robotaxi}, driverless buses~\cite{bus},  and last-mile robotic delivery~\cite{robotruck}. Nowadays, modern autonomous driving systems (ADSs) such as Apollo~\cite{Apollo} make use of machine learning models and logic-based controllers in tandem to process rich sensor data (e.g., images, GPS, point clouds) and plan driving trajectories. 
To ensure the robustness and reliability of ADSs, autonomous driving companies often conduct extensive on-road testing as well as simulation-based testing to cover various driving scenarios. For example, Google Waymo has a fleet of $25000$ virtual vehicles traveling eight million miles per day in simulation~\cite{waymoFleet}. Waymo has also conducted on-road test for more than 20 million  miles in 2021~\cite{waymoSafety}. Despite the significant investment on testing, software bugs still slip through and 
cause
vehicle recalls~\cite{tesla-recall} and traffic accidents~\cite{petrovic2020traffic, waymo_accident, tesla}. 
Therefore, new ADS testing methods are in urgent need to ensure the safety of autonomous vehicles~\cite{lou2021investigation, guo2019safe}.

During ADS development and testing, a common practice is to reuse driving recordings collected from on-road testing or simulation to test new updates to ADSs, such as replacing a traffic light detection model with a new model. This is similar to regression testing in traditional software systems~\cite{yoo2012regression, wong1997study}. However, it is costly to simply replay the entire recordings to test a change, since some recordings include many redundant driving scenes that are not necessary to repetitively test. In particular, a recent study finds that ADS practitioners wish to have tool support to accelerate the testing process and reduce the cost of testing~\cite{lou2021investigation}. 

Test reduction and prioritization techniques have been widely investigated in traditional software engineering~\cite{harrold1993methodology, chen1998new, chen1970heuristics, chen1998simulation, rothermel2001prioritizing, elbaum2000prioritizing, elbaum2001incorporating}. However, these techniques are not directly applicable to ADSs. First, test cases in ADSs are driving recordings with time-series sensor data, rather than discrete program inputs as in traditional software. Therefore, it requires extra care to reduce driving recordings. Second, test coverage metrics~\cite{rothermel2001prioritizing, chen1998new} used in existing techniques are not applicable to multi-module ADSs, since ADSs contain both logic-based code and machine learning models. While many testing techniques have been proposed for ADSs~\cite{zhang2018deeproad, tian2018deeptest, zhou2020deepbillboard, gambi2019generating, abdessalem2018testing, ben2016testing, gambi2019automatically, calo2020generating, li2020av, ebadi2021efficient,abdessalem2018testing2, gambi2019asfault, stocco2020misbehaviour, li2020av, kumar2010genetic, ebadi2021efficient}, most of them focus on test generation rather than test reduction or prioritization. Furthermore, many of them can only handle a single driving model, rather than a multi-module system. 
To the best of our knowledge, only two recent techniques have been proposed to address test reduction or prioritization for ADSs~\cite{birchler2021automated, lu2021search}. However, they can only handle limited driving scenarios, e.g., road shapes only. Furthermore, both of them require access to test configuration files (e.g., the road map) to extract road features, rather than directly from driving recordings. This limits their utility and generalizability in practice. 

To fill the research gap, we propose a novel framework called {\tool} (\underline{S}cenario-based \underline{T}est \underline{R}eduction \underline{a}nd \underline{P}rioritization) that automatically extracts and analyzes rich driving scenarios from driving recordings. In particular, we formally define a driving scene schema with rich features, e.g., pedestrians, traffic lights, stop signs, and interactions. 
We further leverage the publish-subscribe communication mechanism to dynamically extract semantic information related to corresponding features from the communication channels between different modules in an ADS. This publish-subscribe communication mechanism is adopted by most multi-module autonomous driving systems~\cite{Apollo, kato2018autoware} and robot operating system in general~\cite{quigley2009ros}.  Given a driving recording, {\tool} first plays it to extract the corresponding semantic information in each frame of the recording from the communication channels of the ADS under test. The extracted semantic information is encoded to vectors for the ease of comparison and clustering.  
Then, {\tool} slices the driving recording into continuous segments with the same frame vector. Lengthy segments are truncated to reduce the length and redundant segments with the same vector are removed. Furthermore, to expose potential errors (e.g., collisions) early, {\tool} sorts the remaining segments based on the coverage of different driving scene features and the rarity of these features. This heuristic is inspired by coverage-based test prioritization approaches in traditional software engineering~\cite{yoo2012regression}, as well as observations that corner cases or rare driving scenarios are more likely to detect faults~\cite{tian2018deeptest, stocco2020misbehaviour}. 


We implemented {\tool} for an industry-level multi-module ADS called Apollo~\cite{Apollo} and evaluated it in terms of {\em test reduction capability (RQ1)}, {\em test effectiveness of reduced recordings (RQ2)}, and {\em bug detection speed after prioritization (RQ3)}. We created a benchmark of driving recordings on three different types of road maps in a simulator called SVL~\cite{rong2020lgsvl}. We further developed a mutation testing tool to systematically inject errors to different modules of Apollo and evaluated the effectiveness of reduced and prioritized ADS tests. 
The experiment results show that (1) {\tool} reduces over 34\% of given driving recordings and thus significantly reduces the testing time in ADS regressing testing; (2) The reduced driving recordings have comparable test effectiveness---they detected 99\% of injected faults that were detected by the original driving recordings; (3) The diversity-based test prioritization method in {\tool} achieves 39\%, 33\%, and 22\% improvement compared with a coverage-based method related to code change, chronological prioritization, and random prioritization. 

There are three main contributions of this work:
\begin{itemize}
    \item We propose a general scenario-based test reduction and prioritization framework for multi-module autonomous driving systems that adopt the  publish-subscribe communication mechanism.  
    
    \item We make a regression testing benchmark publicly available. The benchmark includes driving recordings on three different kinds of road maps, as well as a mutation testing tool that systematically injects errors in different ADS modules in Apollo and evaluates test effectiveness.\footnote{Our benchmark has been open-sourced in an anonymous GitHub repository at \url{https://github.com/ITSEG-MQ/STRAP}}. With this benchmark, future researchers can systematically evaluate their ADS testing methods in various regression settings.
    
    \item We conduct experiments on an industry-level multi-module ADS and demonstrate the test reduction capability and test effectiveness of our proposed framework.
\end{itemize}

The rest of the paper is organized as follows: Section~\ref{sec:background} introduces the background of current ADS testing practice and multi-module ADSs. Section~\ref{sec:formulation} formulates testing reduction and prioritization problems. Section~\ref{sec:method} describes proposed testing reduction and prioritization methods. Section~\ref{sec:experiment} introduces experiment settings. Section~\ref{sec:result} demonstrates experiment results. Section~\ref{sec:related_work} introduces related works. Section~\ref{sec:threat} describes threats to validity of the work. Section~\ref{sec:conclusion} concludes the paper.

\section{Background}
\label{sec:background}

\begin{figure*}

    \centering
    \includegraphics[width = 0.9\textwidth]{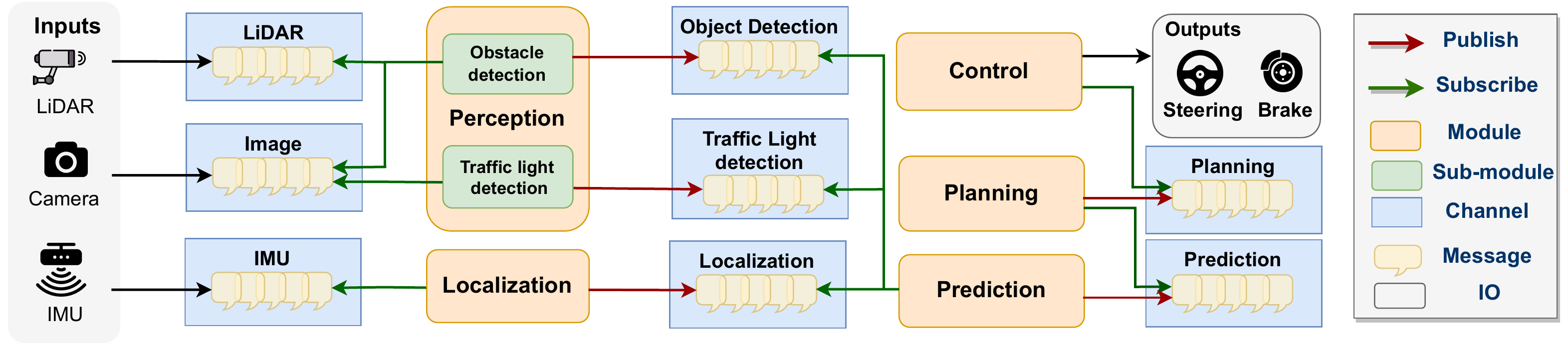}
    \caption{The Architecture and Data Flow of a Multi-Module ADS}
    \label{fig:data_flow}
\end{figure*}

\subsection{ADS Testing Practices in the Industry}


Two primary methods of testing ADSs in the industry are on-road testing and simulation-based testing~\cite{lou2021investigation}. 
The goal is to cover various driving scenarios, especially corner cases, and check consistency and generalizability of ADSs in these scenarios. Such testing processes will produce massive amounts of log data, including sensor data  (e.g., video recordings, point clouds) and outputs of ADS modules (e.g., obstacle detection results, speed control commands). These log data will be further processed for regression testing in a simulation environment. 
When an ADS is updated, it will be tested by replaying previously logged sensor data and comparing the system outputs of the new ADS with previous outputs (e.g., collision or deviation)~\cite{tang2021systematic}. However, replaying an entire recording takes a long time and computation power. Alternative, some ADS developers will manually label and clip recordings to obtain specific driving scenarios (e.g., overtaking, merging) for testing. Yet since this requires a lot of manual effort, it cannot be done in a large scale.


\subsection{Multi-Module ADSs}



Modern ADSs such as Baidu Apollo~\cite{Apollo} and Autoware~\cite{kato2018autoware} are composed of multiple modules to process rich sensor data and make trajectory plans. A typical multi-module ADS includes perception, localization, prediction, planning, and control modules, as shown in Figure~\ref{fig:data_flow}. Similar to robot operating systems (ROS)~\cite{quigley2009ros}, these modules in an ADS communicate and coordinate with each other through {\em the publish-subscribe communication mechanism}. Each module maintains its channel(s) to publish module outputs. Specifically, when a module generates its output at a timestamp, it packages the output as a message and publishes the message into its channel. Each module also subscribes to one or more channels of other modules to obtain needed data.  

Figure~\ref{fig:data_flow} gives an overview of an multi-module ADS based on the architecture of Apollo. As the autonomous vehicle is driving, sensors such as cameras and LiDARs continuously capture environment information, package collected data to messages, and send messages into the corresponding channels. In the perception module, a traffic light detection model subscribes to the message from the image channel and detects traffic lights in the images. The model predictions are then published to the traffic light detection channel. Similarly, the obstacle detection model subscribes to both the image channel and point cloud channel to detect and classify obstacles on the road, such as vehicles and traffic cones. The detection results are published to the obstacle detection channel. The prediction module subscribes data from localization, traffic light detection and obstacle detection channels to predict the trajectories of detected obstacles. The prediction result is published to the prediction channel, which is subscribed by the planning module. Finally, the planning module generates the trajectory of the ego-vehicle, while the control module subscribes data from the planning channel and outputs the control commands such as steering angles and brakes.

\subsection{Regression Testing on Driving Recordings}



\label{sec:recording}

\begin{figure*}
    
    \centering

    \includegraphics[width = 0.9\textwidth]{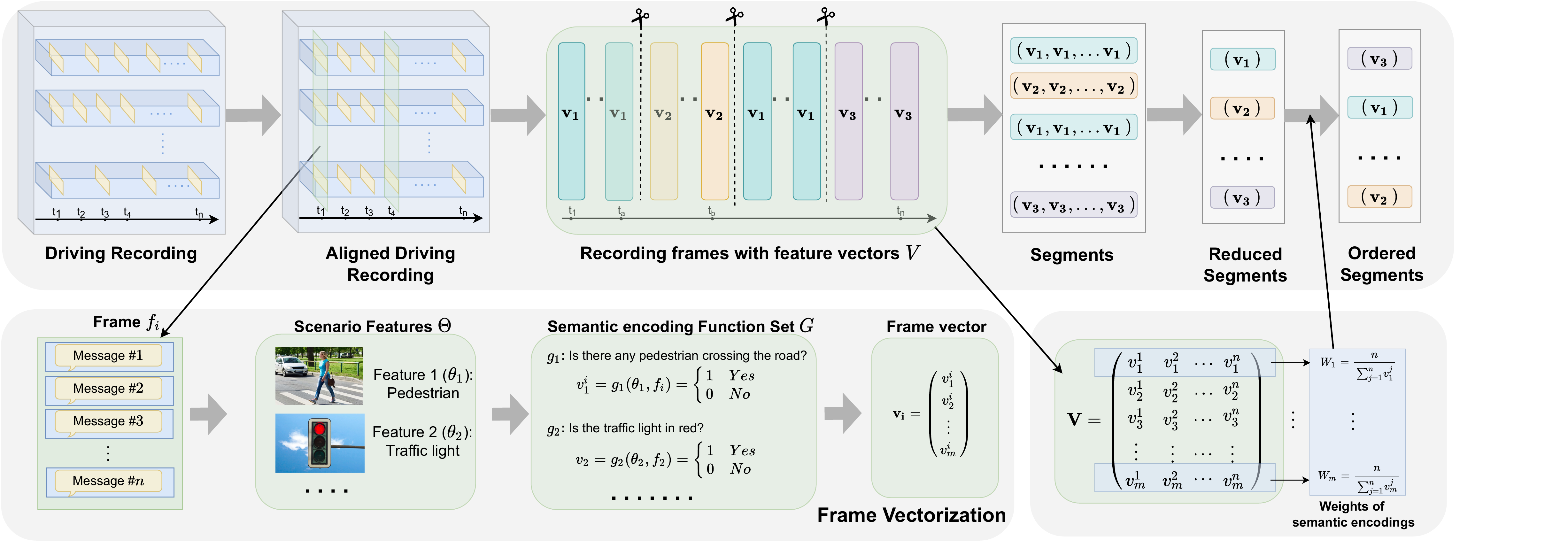}
    \caption{The architecture of proposed regression testing reduction and prioritization method}
    \label{fig:overview}
\end{figure*}

A driving recording includes all sensor data and module outputs that are packaged as messages in different channels during the running of an ADS, as shown in Figure~\ref{fig:overview}.
Mathematically, given a start time point $t_a$ and an end time point $t_b$, a \textbf{driving recording} is denoted as $R_{a \rightarrow b}$. Suppose it contains $n$ channels $\{C_{a \rightarrow b}^i|1 \leq i \leq n]\}$. Each channel $C_{a \rightarrow b}^i$ contains all messages $m_j^i$ created in the time period $[t_a, t_b]$, denoted as $C_{a \rightarrow b}^{i} = \{m_j^i| a \leq j \leq b\}$
where $t_j$ is the timestamp when a message is created. Given a timestamp $t_a$, a \textbf{frame} $f_{a}$ is a slice of a recording that contains all channel messages created at timestamp $t_a$, denoted as $f_a=\{m_a^i|1 \leq i \leq n\}$. Therefore, a driving recording can be seen as a list of frames, denoted as $R_{a \rightarrow b} = \{f_i| a \leq i \leq b\}$.  A \textbf{recording segment} $s_{a' \rightarrow b'}$ is a clip of the driving recording containing frames created in $[t_{a'}, t_{b'}]$, where $a \leq a' \leq b' \leq b$. However, since different modules are invoked in different time orders and frequency, the number of messages and the creation time of messages in different channels are not quite aligned. This message alignment problem is discussed and solved in Section~\ref{sec:alignment}. Table~\ref{tab:vec} summaries terminologies and notations in this paper.

Messages in a channel can be used to perform regression testing on the module that subscribes to the channel. For example, suppose a driving recording $R_{a \rightarrow b}$ with $m$ frames, the image channel $c_{a \rightarrow b}^{img}$ in $R_{a \rightarrow b}$ with $m$ messages of collected images, and the traffic light detection channel $c_{a \rightarrow b}^{light}$ with $m$ messages of traffic light detection results. When the traffic light detection system is updated, by replaying the driving recording, the new version of traffic light detection system can take images from the recording frames and output detection results. Then, by comparing the new detection results with the original detection results in the traffic light detection channel $c_{a \rightarrow b}^{light}$, we can check whether inconsistent model predictions or discrepancies are introduced. For example, if a traffic light detected as in red by the old traffic light detection system but it is detected as in green by the new one, it implies a potential fault. Other ADS modules can be tested in the same way.

\begin{table}
\centering
\caption{Terminology Definitions}
\label{tab:vec}
\scalebox{0.85}{
\begin{tabular}{lll} 
\hline
\rowcolor[rgb]{0.663,0.663,0.663} \textbf{Terminology} & \textbf{Symbol}                 & \multicolumn{1}{c}{\textbf{Meaning}}                                                                                                                                            \\
Recording~                                             & $R_{a \rightarrow b}$            & \begin{tabular}[c]{@{}l@{}}A log file that~stores sensor data and \\module outputs in different channels\\from the timestamp~$t_a$ to $t_b$.\end{tabular}                       \\
\rowcolor[rgb]{0.898,0.898,0.898} Channel~             & $c$                             & \begin{tabular}[c]{@{}>{\cellcolor[rgb]{0.898,0.898,0.898}}l@{}}A message queue to store the outputs\\of a module~ ~ ~ ~ ~ ~ ~ ~ ~ ~ ~ ~ ~ ~ ~ ~ ~ ~ ~ ~ ~ ~ ~ ~~\end{tabular}  \\
Message~                                               & $m_a$                           & \begin{tabular}[c]{@{}l@{}}The output of a module ~created at\\ the timestamp $t_a$\end{tabular}                                                                                \\
\rowcolor[rgb]{0.898,0.898,0.898} Frame~               & $f_a$                           & A slice of the recording at timestamp $t_a$                                                                                                                                \\
Segment~                                               & $s_{a \rightarrow b}$ | $s_i$            & \begin{tabular}[c]{@{}l@{}}A list of frames created from  $t_a$  to $t_b$ | \\ The $i^{th}$ segment in a segment list           \end{tabular}                                                                     \\
\rowcolor[rgb]{0.898,0.898,0.898} Frame feature~       & $\theta_i$                      & \begin{tabular}[c]{@{}>{\cellcolor[rgb]{0.898,0.898,0.898}}l@{}}An attribute to represent specific \\driving scenario-related~ semantic\\information in a frame\end{tabular}    \\
Frame feature value~                                   & $v_i^a$                         & \begin{tabular}[c]{@{}l@{}}The value of the frame feature $\theta_i$ for\\~the frame $f_a$\end{tabular}                                                                         \\
\rowcolor[rgb]{0.898,0.898,0.898} Frame vector~        & $\mathbf{v}_a$                  & \begin{tabular}[c]{@{}>{\cellcolor[rgb]{0.898,0.898,0.898}}l@{}}A vector $(v_1^a, v_2^a, ..., v_n^a)$ to represent \\~the frame $f_a$\end{tabular}                              \\
Segment vector~                                        & $\mathbf{sv}_{i}$ & A vector to represent the segment $s_{i}$                                                                                                                        \\
\hline
\end{tabular}}
\end{table}

\section{Problem Formulation}
\label{sec:formulation}
%


In this section, we formulate the test reduction problem and the test prioritization problem in ADS testing.

\textbf{Test reduction:} Given a driving recording $R_{a \rightarrow b}$, a test reduction method $\alpha$ should output a list of recording segments $S = \{s_1, s_2, ..., s_n\}$, denoted as $S = \alpha(R_{a \rightarrow b})$, where $n$ is the number of recording segments. The total time cost of replaying recording segments in $S$ should be less than the time cost of replaying $R_{a \rightarrow b}$. Meanwhile, $S$ should be able to detect the same number of faults as $R_{a \rightarrow b}$. More specifically, test reduction shall comply with two constraints:  $\sum_{i=1}^{n}(T(s_i)) \leq T(R_{a \rightarrow b})$ and  $\sum_{i=1}^{n}(\tau(s_i, SUT)) = \tau(R_{A \rightarrow b}, SUT)$, where $n$ is the number of recording segments, $T$ is a function to output replaying time cost of a given segment or recording, $SUT$ is the testing module, and $\tau$ represents the number of detected faults in the testing module using the input segment or recording.




\textbf{Test prioritization:} Given $n$ recording segments $S$$=$$\{s_1,s_2,...s_n\}$ that detect $m$ faults in the new version of an ADS module, test prioritization aims to sort the recording segments and obtain a list $P = [p_1, p_2, ..., p_n]$, where $p_i$ is the execution order of the recording segment $s_i$. For example, if $p_1 = 5$, the recording segment $s_1$ will be the fifth to replay.  The sorted recording segments should detect all $m$ faults as early as possible.

\section{Approach}
\label{sec:method}

Figure~\ref{fig:overview} shows the overview of our approach, {\tool}. First, given a driving recording, {\tool} performs message alignment across different channels and converts messages in each time frame into vectors based on a driving scene schema (Section~\ref{sec:record2vec}). 
Second, {\tool} slices the given recording into segments based on the similarity of consecutive vectors. Long segments of the same continuous vectors will be truncated and redundant segments will be removed (Section~\ref{sec:reduction}). Third, the remaining segments are prioritized based on the coverage of driving scene features and the rarity of these features (Section~\ref{sec:prioritization}).

\subsection{Recording Alignment and Vectorization}
\label{sec:record2vec}

\subsubsection{Recording message alignment}
\label{sec:alignment}
As different modules in ADS run asynchronously and in different frequencies, the timestamp and the number of messages in different channels are not aligned. For example, in Apollo, the localization module logs a message every 0.1 second, while the prediction module logs a message every 0.15 second. Since the messages in different channels are not aligned, raw driving recordings cannot be used as-is for comparison and clustering. Therefore, we need to align them first. 

\begin{figure}

    \centering

    \includegraphics[width = 0.45\textwidth]{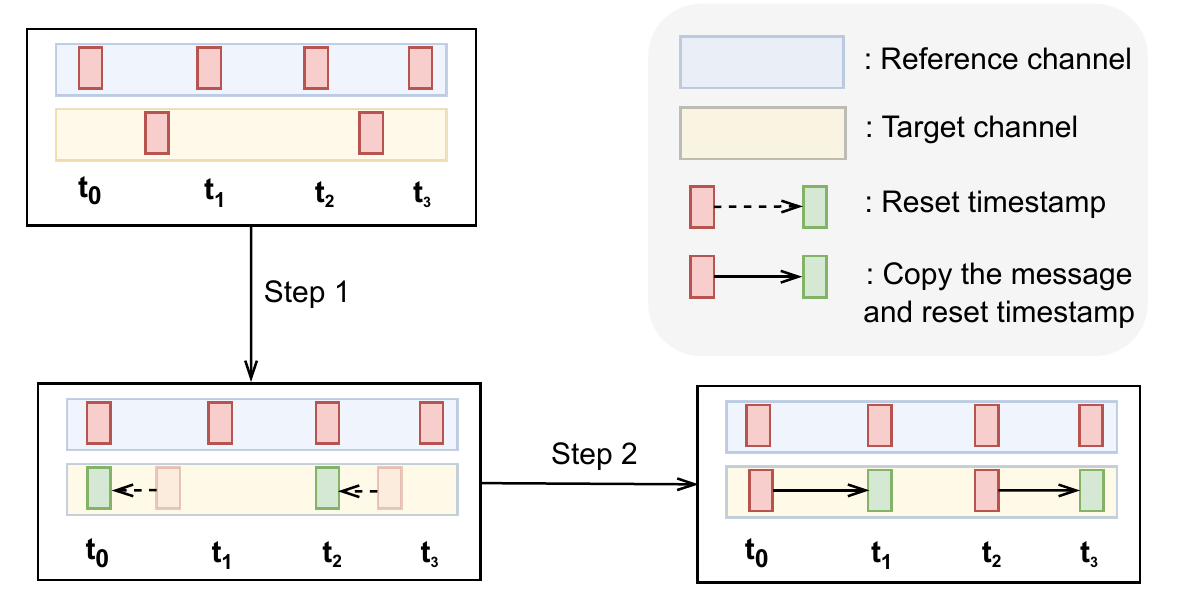}
    \caption{An example of message alignment}
    \label{fig:alignment}
\end{figure}

We propose a novel message alignment algorithm to address this issue. Specifically, we choose the channel containing most messages as the reference channel to align messages of other channels (i.e. target channels). Given the reference channel $C_{ref}$, at each time we retrieve two consecutive messages $m_{ref}^i$ and $m_{ref}^{i+1}$ from it and obtain their creation timestamps $t_i$ and $t_{i+1}$. Then, for the target channel, if the message is generated between $(t_i, t_{i+1})$, we reset the timestamp of the message to $t_i$, as shown in Step $1$ of Figure~\ref{fig:alignment} . Then, we iterate over the reference channel and the target channel to check whether at $t_i$ there is a corresponding message in the target channel. If no message is found, we copy the prior message at $t_{i-1}$ in the target channel and reset its timestamp to $t_i$ as shown in Step $2$ of Figure~\ref{fig:alignment}.  In this way, we align all channels to ensure that the aligned driving recording can be sliced to a list of frames.

\begin{figure}[t]
\centering
{\small
$\def\arraystretch{1.1}
\setlength{\arraycolsep}{1pt}
\begin{array}{rcl}
scene & := & \epsilon\ |\ object \ \mbox{;}\ scene \\
object & := & dynamic\ object |\ static\ object  \\
dynamic\ object & := & <actor,\ action>\ \\
actor & := & \mbox{\ttt vehicle}\ |\ \mbox{\ttt pedestrian}\ |\ cyclist\ |\ \mbox{\ttt unknown} \\
vehicle & := & \mbox{\ttt truck}\ |\ \mbox{\ttt car}\ |\ \mbox{\ttt bus}\ |\ \mbox{\ttt van} \\
cyclist & := & \mbox{\ttt bicyclist}\ |\ \mbox{\ttt motorcyclist}\ |\ \mbox{\ttt tricyclist}\ \\
action & := &  \mbox{\ttt stop}\ |\ \mbox{\ttt cruise}\ |\ \mbox{\ttt change lane}\ |\ \\
& & \mbox{\ttt overtake}\ |\ \mbox{\ttt cross}\ |\ ... \\
static\ object & := & \mbox{\ttt traffic\ light}\ |\  \mbox{\ttt stop sign}\ |\ \mbox{\ttt crosswalk}\ |\  \\
& & \mbox{\ttt intersection}\ |\ \mbox{\ttt traffic\ cone}\ |\ ...\ |\ \mbox{\ttt unknown} \\
traffic\ light & := & <color, shape, orientation> \\
color & := & \mbox{\ttt red}\ |\ \mbox{\ttt green}\ |\
\mbox{\ttt yellow}\ |\ \mbox{\ttt black} \\ 
shape & := & \mbox{\ttt square}\ |\ \mbox{\ttt round}\ \\
orientation & := & \mbox{\ttt vertical}\ |\ \mbox{\ttt horizontal}\ \\
\end{array}$
}
\caption{Part of the Driving Scene Schema}
\label{fig:grammar}
\end{figure}

\subsubsection{Schema-based Recording Vectorization}
Given an aligned driving recording, {\tool} converts each frame into a vector $\mathbf{v}$, where each dimension represents specific semantic information in a driving scene. 
We formally define a schema for a variety of semantic information in a driving scene, as shown in Figure~\ref{fig:grammar}. A driving scene is represented as a list of static or dynamic objects. A dynamic object is defined of an actor (e.g., vehicles, pedestrians, etc.) and an action of the actor (e.g., stop, cruise, etc.). A static object can be traffic light, stop sign, crosswalk, interaction, traffic cone, etc. An object can have subcategories and also properties. Our driving scene schema is designed as a general schema for all driving systems. Yet one can customize it based on the concrete channel messages logged by an ADS. For example, in Apollo, each detected object is also logged with its coordinates and heading direction (if movable). Such information can also be encoded as new dimensions in the vector. The encoding function in our current implementation only considers the existence of different types of objects and their properties in a driving scene. Given channel messages in a time frame, it parses the channel messages into a list of objects based on the schema and flattens the list of extracted objects and their properties into a vector of integers using label encoding. Specifically, 0 is a reserved code for {\ttt none} (i.e., undetected objects or properties). This encoding function also enforces a specific ordering of objects based on their types for the ease of vector comparison in the next step. For example, consider a driving scene that contains  vertically-aligned, round traffic lights with the red light on. The corresponding 
feature dimensions in the final vector is <$22, 34, 39, 41$> where 22 is the unique code for traffic light, 34 is the code for red, 39 is the code for the round shape, and 41 is the code for vertical alignment. If no traffic light is detected, the four dimensions are all set to 0.

After vectorization, a driving recording is represented as a list of vectors, where each vector corresponds to each time frame in the recording. In regression testing, if one or more updated ADS modules are given as input, {\tool} will further simplify the vectors by only retaining features obtained from the channels that the updated modules subscribe or publish to. For example, suppose only the traffic light detection model is updated. Since the traffic light detection model takes raw images as input and publishes prediction results to the traffic light detection channel, features obtained from the traffic light channel (e.g., traffic light color, shape, and orientation) will be retained in the final vectors. To avoid over-simplifying the vectors, we specify several features that should always be preserved in the final vectors, including stop sign, intersection, and crosswalk.

\subsection{Test Reduction}
\label{sec:reduction}

\subsubsection{Segmentation}
\label{sec:segmentation}

\begin{algorithm}
\SetAlgoLined
\SetKwInput{Input}{Input~}
\SetKwInput{Output}{Output~}
    \SetKwInOut{Initialize}{initializiton}
    \SetKwData{Count}{count} \SetKwData{Weight}{weights} \SetKwData{Rare}{rarity} \SetKwData{Pm}{p}
    \SetKwFunction{Argsort}{Argsort}
    \SetKwFunction{Append}{append}
    \SetKwFunction{Maj}{Majority}
    \SetKwData{Vector}{V}
    \SetKwData{l}{l}
    \SetKwData{w}{w}
    \SetKwData{Left}{left}\SetKwData{This}{this}\SetKwData{Up}{up} \SetKwData{Sstart}{ss} \SetKwData{Ds}{S$_{temp}$}
    \SetKwData{Dsr}{S} \SetKwData{Svr}{SV}
    \SetKwData{Unique}{uni\_vec} \SetKwData{Sv}{SV$_{temp}$} \SetKwData{Window}{window} \SetKwData{Vsm}{V$_{temp}$} \SetKwData{Send}{se}
    \SetKwFunction{GSegment}{GetSegment} \SetKwFunction{Clip}{Clip}
    \SetKwFunction{Contain}{contain}
 \Input{\Vector:  a list of frame feature vectors with size \(N\), $N$ is the number of frames; \\
\ \ \ \ \ \ \ \ \ \ \ \ \ \ \l: the clip length of a segment; \\
\ \ \ \ \ \ \ \ \ \ \ \ \  \w: the sliding window size; \\
 }
 
 \Output{\Dsr: a list of reduced driving segments; \\
 \ \ \ \ \ \ \ \ \ \ \ \ \ \ \ \  \Svr: a list of segment vectors;}
 
 \Ds, \Sv $\leftarrow$ []\;
 \ccomment{\tcp{Initiate the start and end indices of a segment}}
 \Sstart, \Send  $\leftarrow$ $0$;

 \ccomment{\tcp{Create an empty list to save smoothed feature vectors}}
\Vsm $\leftarrow$ [];

\ccomment{\tcp{Smoothing using sliding window}}
\For{$i \leftarrow 1$ \KwTo $N$}{
    \ccomment{\tcp{Obtain the window of frame feature vectors}}
    \eIf{$i \leq \frac{w}{2}$}{
        \Window $\leftarrow$ \Vector[$i:i+w-1$]\;
    }{
      \eIf{$i \geq N-\frac{w}{2}$}{
        \Window $\leftarrow$ \Vector[$i:i+w-1$]\;
      }{
        \Window $\leftarrow$ \Vector[$i-\frac{w}{2}:i+\frac{w}{2}$]\;
      }
    }
     \ccomment{\tcp{Set V$_{temp}$[$i$] as the majority vector in the window}}
    \Vsm.\Append{\Maj{\Window}};
 }
 
 \ccomment{\tcp{Recording segmentation}}
	 \For{$i\leftarrow 2$ \KwTo $N$}{ 
	  \eIf{\Vector[$i$] $\neq$ \Vector[$i-1$] or $i$ == $N$}{
	  \Send $\leftarrow$ i \;
	  }{
	    \Ds.\Append{\Vsm$[\Sstart:\Send]$}\;
	    \Sv.\Append{\Vsm$[i-1]$}\; 
	     \ccomment{\tcp{Reset segment indices}}
	    \Sstart, \Send $\leftarrow i$;
	  }
    }

\ccomment{\tcp{Clip each segment to length $l$}}
  \ForEach{$s$ in \Ds}{
     $s \leftarrow $ \Clip{$s$, $l$}; 
  }
  
\ccomment{\tcp{Drop segments with the same segment vector}}  

\Dsr, \Svr $\leftarrow$ []\;
\For{$i \leftarrow 1$ \KwTo \Ds.length}{
    \If{!\Svr.\Contain{\Sv$[i]$}}{
        \Svr.\Append{\Sv$[i]$}\;
        \Dsr.\Append{\Ds$[i]$}\;
    }
}

 \Return \Dsr, \Svr\;

 \caption{Recording Segmentation and Reduction}
 \label{alg:reduction}
\end{algorithm}

 \begin{figure}
     \centering
     \includegraphics[width=0.4\textwidth]{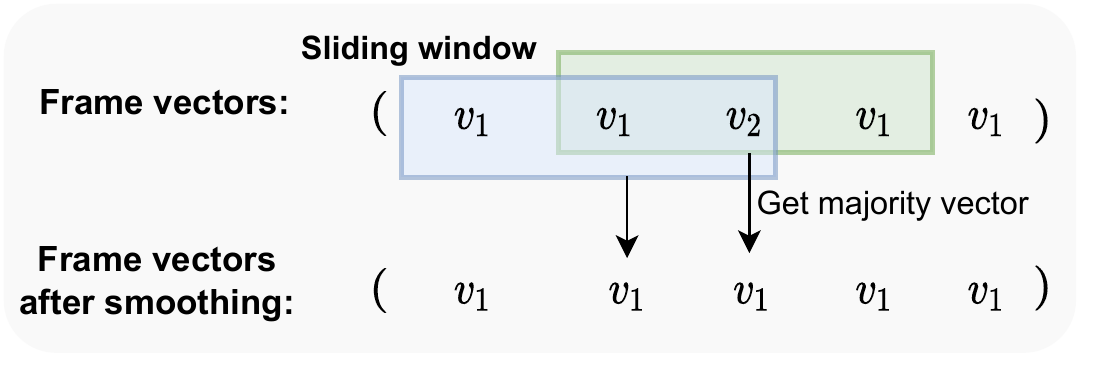}
     \caption{Smoothing frame vectors with a sliding window}
     \label{fig:window}
 \end{figure}
 
{\tool} slices the given recording into segments based on the similarity of consecutive vectors. If vectors in a consecutive time range are the same, then they will be sliced into one single segment. For example, in a 1-hour driving recording, if the ego-vehicle drives on a highway with no new objects detected for 10 minutes, the vectors for each time frame in this 10 minutes will be the same and therefore this 10 minutes will be sliced into a single segment. 

In practice, we noticed that due to noises in raw sensor data and the uncertainty of DL models, channel messages often contain {\em glitches}---exceptional predictions that only exist for a very short period of time (e.g., 0.1 second). Such glitches often do not lead to abnormal ADS behavior since they only exist very shortly and ADS make decisions based on a sequence of time frames, not just one. However, it has a significant impact on our segmentation algorithm. Specifically, it will lead to many tiny segments since a glitch will produce an exceptional vector that is different from its preceding and following vectors. To handle this issue, {\tool} smooths glitches in the list of vectors using a sliding window, as shown in Lines $4-16$ of Algorithm~\ref{alg:reduction}.
Figure~\ref{fig:window} illustrates this process with a sliding window of 3 vectors ($n=3$). When smoothing is finished, we split frames into segments. Each segment contains a list of frames represented by the same vector, which is denoted as the segment vector.

\subsubsection{Segment Reduction}
{\tool} adopts two strategies to reduce the length and the number of segments. 
First, for each driving segment, 
as each frame in the segment is semantically equivalent (i.e., represented by the same vector),
{\tool} only keeps the first \(n\) frames in the segment. {\tool} keeps n frames rather than one frame, because certain ADS modules, such as the planning module, rely on a sequence of frames to make a decision, not just one frame. In our current implementation for Apollo, we experimented with different number of frames and finally set n to 45 (roughly 3 seconds). 
Second, as each segment reflects a unique driving scenario, we only keep one unique segment and remove those segments with identical segment vectors. By combining these test reduction strategies, we obtain a list of reduced segments as test cases to evaluate ADS modules.

  

\subsection{Test Prioritization}
\label{sec:prioritization}








\begin{algorithm}
\SetAlgoLined
\SetKwInput{Input}{Input~}
\SetKwInput{Output}{Output~}
    \SetKwInOut{Initialize}{initializiton}
    \SetKwData{Count}{count} \SetKwData{Weight}{weights}
    \SetKwData{Fv}{V} \SetKwData{S}{s} \SetKwData{Sv}{SV}
    \SetKwData{Rare}{rarity} \SetKwData{Pm}{p}
    \SetKwFunction{Argsort}{DescendingArgsort}
    \SetKwFunction{Nonzero}{Nonzero}
 \Input{\Fv: a list of frame feature vectors with size \(N \times Q \); $N$ is the number of frames, $Q$ is the number of elements in a frame feature vector; \\
 \ \ \ \ \ \ \ \ \ \ \ \ \ \Sv: a list of segment vectors with size \(M \times Q\); $M$ is \\
 \ \ \ \ \ \ \ \ \ \ \ \ \ the number of segments after test reduction; \\
 }

 \Output {\Pm: a list of segments' execution order with size \(M\);}
 
 \Weight $\leftarrow$ []\;
\ccomment{ \tcc{Initialize a list \Weight to save the weights of feature elements and a list \Rare to save the rarity degree of segments}}
 \For{$i \leftarrow 1$ \KwTo $Q$}{
    \Weight[$i$] $\leftarrow$ 0\;
    \Rare[$i$] $\leftarrow$ 0\;
 }
 
 \ccomment{\tcp{Calculate weights of feature elements}}
 \For{$i \leftarrow 1$ \KwTo $Q$}{
    \Weight[$i$] $\leftarrow \frac{N}{\sum_{j=1}^{N}\Nonzero{V[i][j]}}$ \;
 }
 
 \ccomment{\tcp{Weight normalization}}
  	 \For{$i\leftarrow 1$ \KwTo $Q$}{
 	    \Weight[$i$] \(\leftarrow \tfrac{\Weight[i]}{\sum_{j=1}^{Q}\Weight[j]}\) \;
 	 }
 
 \ccomment{\tcp{Calculate rarity of each segment}}
 \For{$i \leftarrow 1$ \KwTo $M$}{
  \Rare[$i$]  $ \leftarrow \sum_{j=1}^{Q}(\Weight[j] \times SV[i][j])$\;
 }
 
 \ccomment{\tcp{Get indices of \Rare after descending sorting}}
 \Pm $\leftarrow$ \Argsort(\Rare)\;
 
 \Return \Pm\;
 
 \caption{Coverage and Rarity Based Prioritization}
 \label{alg:prioritization}
\end{algorithm}

Coverage-based prioritization methods have been widely investigated in traditional software systems. These methods sort test cases based on the total or additional coverage of program statements or branches~\cite{aggrawal2004code, yoo2012regression}. Inspired by these methods, we propose the notion of {\em semantic coverage} of different driving scenes by counting the number of non-zero feature dimensions in a vector. Note that 0 is a preserved code that indicates the corresponding object or property does not exist. Intuitively, a driving scene with a traffic light and a pedestrian in a crosswalk has a higher coverage than a driving scene with only a pedestrian. However, only considering the coverage of driving scenes may not be sufficient. Since ADSs are equipped with DL models for perception and prediction, common driving scenes are less likely to expose faults compared with rare driving scenes or corner cases, since common driving scenes are prevalent in training data. This phenomenon has been observed by several existing works in ADS testing~\cite{tian2018deeptest, stocco2020misbehaviour}. Therefore, we should also take into account the rarity of driving scenes when prioritizing recording segments. 

We define a new prioritization method that accounts for both the coverage and rarity of different driving scenes. First, given a driving recording represented by a list of frame vectors $R=\{v_1, v_2, ..., v_n\}$, we calculate the rarity score of each dimension in the vector by the formula at Line $9$ of Algorithm~\ref{alg:prioritization}, where $n$ is the number of frames, {\ttt{nonzero($V[i][j]$)}} $= 1$ when $V[i][j] \neq 0$. The formula presents that the semantic information occurring less times in a driving recording has a higher rarity score. The process is shown in Lines $8-10$ of Algorithm~\ref{alg:prioritization}. Then, we normalize weights to $[0,1]$ (Lines $12-14$ of Algorithm~\ref{alg:prioritization}).   
Then, for each recording segment represented by its segment vector, it sums the rarity score of non-zero feature elements in the corresponding segment vector (Lines $16-18$ of Algorithm~\ref{alg:prioritization}). Finally, we sort the list of recording segments in descending order based on the rareness of recording segments. The sorted indices of the recording segments list is obtained as the execution orders of recording segments.

\section{Experiments}
\label{sec:experiment}
\subsection{Research Questions}
To evaluate the effectiveness of {\tool}, we conducted experiments to answer the following research questions:

\begin{itemize}

\item RQ1: To what extent can the proposed test reduction method reduce a given driving recording?

\item RQ2: Compared with an original recording, how effective is the reduced recording in detecting faults?

\item RQ3: Compared with alternative prioritization methods, how effective is the proposed test prioritization method in detecting faults?

\end{itemize}

We implemented and evaluated {\tool} on an industry-level ADS, Apollo 5.0~\cite{Apollo}. We chose Apollo 5.0 since it is a mature version with stable modules. We constructed our experiment as follows.
First, we collected driving recordings from a simulator SVL~\cite{rong2020lgsvl}. Then, we injected mutants into the source code of Apollo to simulate the module changes with faults. 
Third, we tested the traffic light detection module, the obstacle detection module, the planning module and the prediction module to build baselines.
Finally, we applied {\tool} 
to split driving recordings, generated reduced and prioritized test segments, and evaluated {\tool}'s effectiveness.


\subsection{Benchmark Creation}
\begin{figure}
    \centering
    \includegraphics[width = 0.45\textwidth]{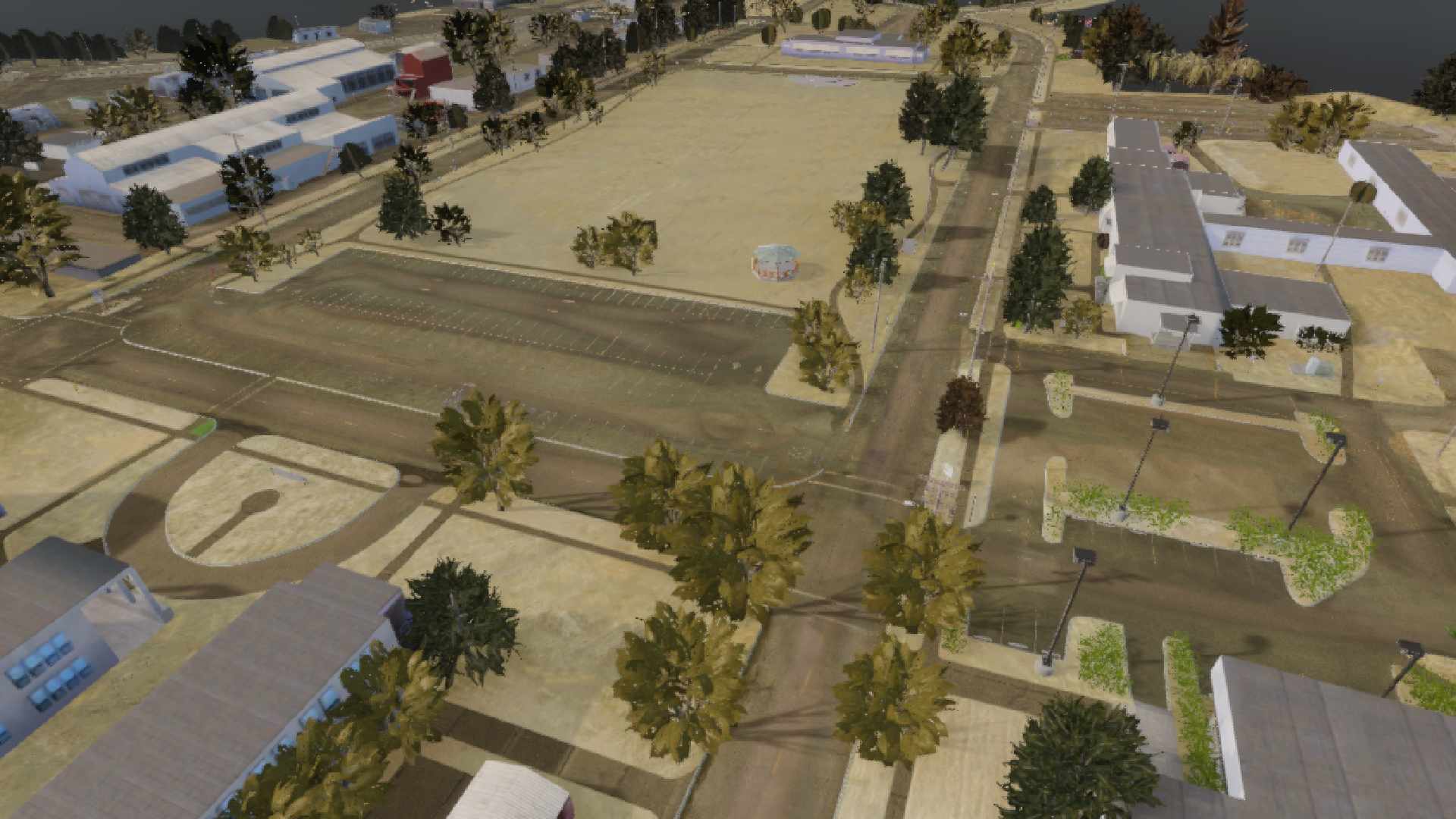}
    \caption{The Gomentum map in SVL}
    \label{fig:gomentum}
\end{figure}

We created simulation environments using SVL~\cite{rong2020lgsvl}. SVL is a high fidelity simulator to render driving environments, which provides bridges to connect with ADSs such as Apollo and Autoware. The rendered driving environment is fed to an ADS via the bridge. Then, the ADS outputs control commands and sends them back to SVL to render the next driving scene based on the control commands. We collected driving recordings from three maps including Cubetown, Gomentum, and Shalun in SVL to create three testing suites. Specifically, Cubetown is a simple map containing a circular road. Gomentum (as shown in Figure~\ref{fig:gomentum}) and Shalun are reconstructed maps based on real-world autonomous vehicle testing (e.g., on-road testing) environments. These three maps cover the various driving environments in urban and rural areas. The weather condition and non-player characters (e.g., vehicles and pedestrians) behaviors are randomly generated by the simulator in default settings. For each test suite, the destination waypoints are randomly generated by the simulator. To collect sufficient data, we manually inspected the waypoints and chose those waypoints with the most complex trajectories. When Apollo 5.0 was connected with SVL, we ran a recorder in Apollo to record all channel messages generated during the simulation.   In the end, collected test suites contain driving recordings of different lengths (Cubetown: 84.14 seconds, Gomentum: 96.22 seconds, and Shalun 73.28 seconds). For each driving recording, we applied proposed reduction method to obtain reduced driving segments. Each driving segment is separately replayed to test the ADS. To ensure that Apollo is in a correct state before replaying each segment, {\tool} replayed the one second before the segment in the original recording to initialize modules and restore system states.




\begin{figure}
    \centering
    \includegraphics[width = 0.45\textwidth]{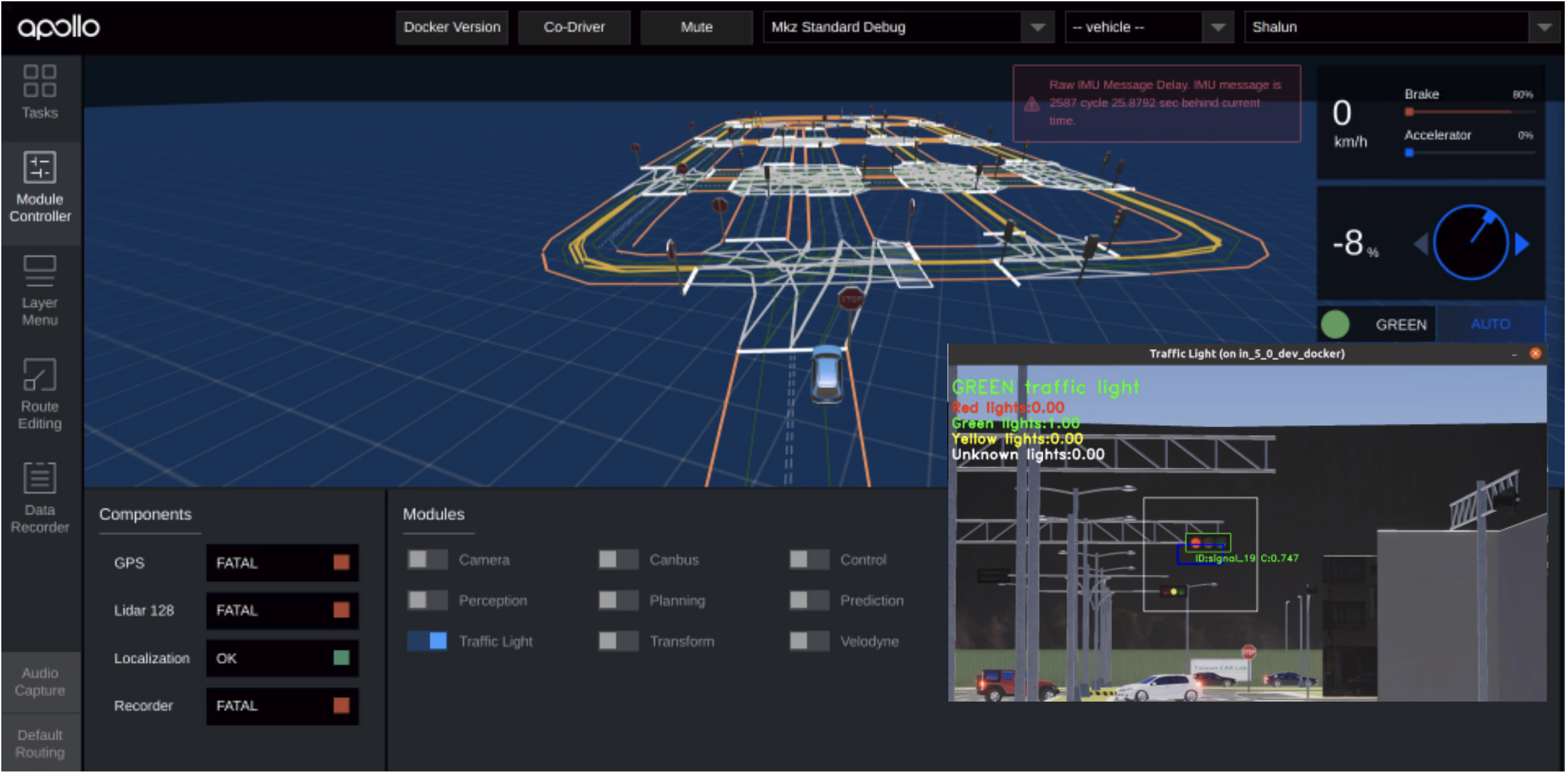}
    \caption{A bug detected by replaying the recording segment. A new traffic light detection model classifies a red light as a green light, which is inconsistent with the previous run}
    \label{fig:bug}
\end{figure}

\begin{figure}[htp]

\subfloat[A real fault of correcting buffer size]{%
  \includegraphics[clip,width=\columnwidth]{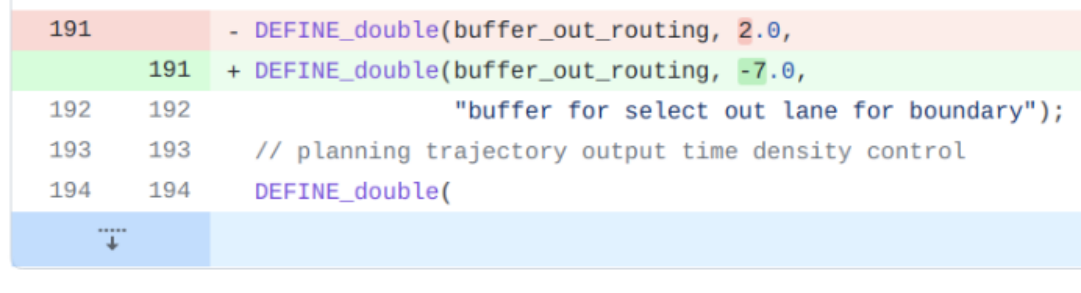}%
}

\subfloat[A mutant of changing constant value]{%
  \includegraphics[clip,width=\columnwidth]{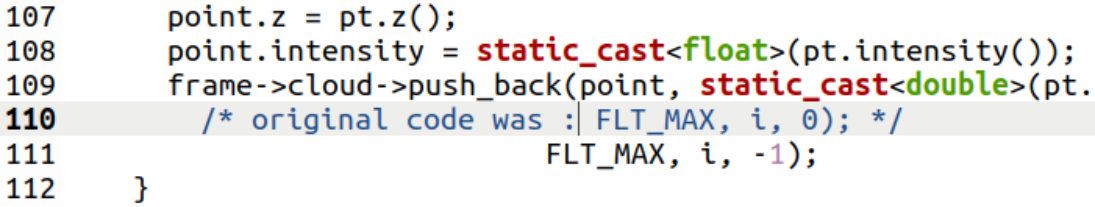}%
}

\caption{An example of real and artificial fault.}
\label{fig:mutant_bug}
\end{figure}

We modified the source code of four modules to simulate the module change with faults. For those modules under test, we implemented a mutation testing~\cite{jia2010analysis} tool to randomly inject mutants into the source code. The mutants include commonly used mutation operators such as replacing arithmetic operators, changing constant values, changing variable values, and changing condition operators, where the generated mutants can simulate real faults
\footnote{Changing  constant values:\url{https://bit.ly/3Qsv7Cn}}\footnote{Changing condition statement: \url{https://bit.ly/3w13194}}\footnote{Replacing arithmetic operators: \url{https://bit.ly/3vZc2zG}}. An example\footnote{\url{https://bit.ly/3AgZyWz}} is shown in Figure~\ref{fig:mutant_bug}. For each module, $9$ c++ files were randomly selected for mutation and totally, $36$ regression testing benchmarks are created for each test suite.


\subsection{Evaluation Metrics}
For RQ1, we used the \textbf{reduction ratio of testing time} to reflect the effectiveness of the test reduction method, which is defined as the length of the reduced driving recording over the length of the original driving recording.







For RQ2, we used the \textbf{fault coverage} to reflect the quality of the test reduction method, which is defined as the ratio of the number of the detected faults in reduced driving recording over the number of the detected faults in the original driving recording. In our work, we defined a fault as the inconsistency of a module output before and after code changes (e.g., injecting a mutant). Since each output is a driving segment, we applied the frame vectorization approach to such output and retrieved the frame vector list representing the output. For the outputs before and after change, we compared each frame pair. Intuitively, if we find any mismatched frame pairs, we claim an inconsistent case is detected. To reduce the systematic noises introduced by Apollo 5.0, we recognized the inconsistency only when more than 10\% of frames were mismatched. This treatment was proposed since our benchmark study showed that the inconsistency rate could be up to 10\% when we replicated experiments on the benchmarks.



We calculated the number of faults from the reduced and the original driving segments. For those detected faults, we manually checked the visualized driving segments (before and after a ADS module change) and removed semantically identical faults (e.g., red traffic light is recognized as green, as shown in Figure~\ref{fig:bug}) to improve experiment quality.

For RQ3, we used metrics \textbf{Top-K} and \textbf{Average Percentage of Faults Detected (APFD)}~\cite{srivastava2008test} to evaluate the effectiveness of test prioritization. Given \(n\) driving segments containing \(m\) bugs, Top-K measures the amount of segments required for finding the first fault. APFD measures the capability of fault detection per percentage of test cases execution. The calculation of APFD is shown in Formula~\ref{eq:apfd}, 

\begin{equation}
\label{eq:apfd}
    APFD = 1 - \frac{\sum_{i=1}^{m}TF_i}{mn} + \frac{1}{2n}
\end{equation}

\noindent where \(n\) is the number of test recording segments, \(m\) is the number of faults, \(TF_i\) is the index of the first segment that detects fault \(i\). For example, if \(TF_2=3\), it means the second fault is detected at the first time when replaying the third prioritized recording segment. A higher APFD value means the test suite can find all faults faster.

\subsection{Experiment Settings}
For test reduction experiments, we clipped each segment by keeping its first $45$ frames as explained in Section~\ref{sec:reduction}. 
To evaluate the test prioritization method, we created three baselines. The first baseline sorts the segments in chronological order (i.e., \textit{CH}). The second baseline randomly sorts the segments (i.e.,  \textit{RD}).
Specifically, we randomly sorted the reduced segments by $100$ times and evaluated the average performances using Top-K and APFD. For the third baseline, we first identified the code change in which function in the source code. Then, we counted the number of calls on the changed function for each recording segment based on the Apollo running log data. Finally, we sorted recording segments in descending order by the number of calls on the changed function (i.e., \textit{CC}). In addition, we marked the proposed semantic coverage method as \textit{SC} and the rarity and semantic coverage-based method as \textit{RSC} in Section~\ref{sec:result}.
All experiments were conducted on a Ubuntu PC with Intel i7-8700 3.2GHz, 32GB of memory, and a NVIDIA GTX 1080Ti GPU.


\section{Results}
\label{sec:result}
\subsection{RQ1: Test Reduction Capability}
\label{sec:result_rq1}
\begin{figure}
    \centering

    \includegraphics[width=0.43\textwidth]{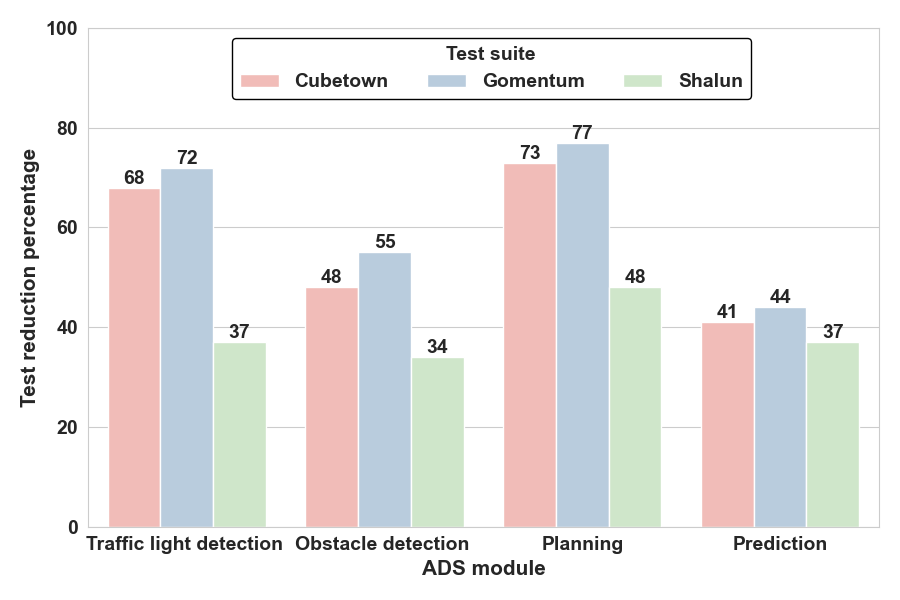}
    \caption{Test reduction percentage when updates are made on different ADS modules}

    \label{fig:exp1}
\end{figure}

    Figure~\ref{fig:exp1} shows the test reduction percentage for three test suits on four testing modules. 
    The result shows that for the planning module in Gomentum, the reduction ratio of testing time achieves $77\%$. Similarly, the reduction ratio of testing time for the planning module in Cubetown achieves $73\%$. 
    A similar pattern can be observed for the traffic light detection and obstacle detection modules in Gomentum and Cubetown where the test reduction ratios are in general close or above $50\%$. 
    
    We noticed the test reduction ratios are relatively lower for modules in Shalun ($34\%$ to $37\%$). We speculated the main reason is that Shalun contains more complex driving scenarios and finer-granularity frame feature encoding functions can be explored besides just using existence of semantic information. We also noticed that test reduction ratios for all test suites on the prediction module are relatively lower than on other modules. 
    We manually inspected the output from the prediction module in Apollo and noticed that the trajectory prediction for vehicles in the ADS tends to fluctuate a lot. For instance, for a vehicle in front of the ego-vehicle, the prediction module in one frame detects the vehicle is driving in a straight line ahead, but in the next frame the prediction result is the vehicle is turning right. Such dynamic prediction changes across frames will result in different semantic feature encoding for continuous frames for the prediction module thus impacting the reduction ratios.  Nevertheless, the reduction ratio is still promising with the minimum reduction ratio of $37\%$ in the prediction module.

\subsection{RQ2 \& RQ3: Test Effectiveness of Test Reduction and Prioritization}

In Table~\ref{tab:exp2_1}, we
present the effectiveness of the reduced recording in detecting faults on the left-hand side. For the traffic light detection system, besides Shalun, in other two test suites, the fault coverage is $100\%$. For Shalun, the fault coverage is $92.3\%$ which is also promising. For the $6$ missing faults, 
we manually inspected the traffic light module output and noticed for vertical traffic light in Shalun (in other two suites, the traffic lights are mostly horizontal), the module output sometimes gave wrong bounding box for traffic lights in a very short duration but the glitch is within our threshold for smoothing ($10\%$). This might contribute to the $6$ missing faults as such short-duration glitch is embedded in those driving frames discarded. The $10\%$ smoothing threshold is empirically retrieved as the best value for Apollo through our pilot study. Overall, the fault coverage achieves on average $98.8\%$, which proves the reduction technique is effective across four modules in different test suites.


We also present on the right-hand side of Table~\ref{tab:exp2_1} the comparison results of our proposed \textit{SC} and \textit{RSC} against the baselines using Top K and APFD metrics.
For each module in a specific test suite, the presented results are the averaged values on nine benchmarks. 
We observed that \textit{RSC} outperforms other baselines for all test suites for prediction module both in Top K and APFD. Similarly, for planning module, \textit{RSC} also outperforms others both in Top K and APFD except in the Shalun test suite for Top K ($1.44$ VS $1.0$ both in \textit{CC} and \textit{SC}).
We speculated that even a specific driving segment containing the rarest semantic information are ranked as the top in \textit{RSC}, this driving segment does not contain faults. However, empirically the correlation between semantic rareness and faults is very positive as shown in the table as a general trend.  
For obstacle module, \textit{RSC} has lower value in APFD than the baselines ($0.47$ VS $0.51$ in \textit{CH}) in Gomentum. 
Similarly, in traffic light detection system, \textit{RSC} outperforms all other baselines in Top K, but for APFD has lower value both in Gomentum ($0.76$ VS $0.82$ in \textit{SC}) and Shalun ($0.54$ VS $0.56$ in \textit{SC}).
We believe because the semantic information in both traffic light and obstacle detection modules are largely static. Thus capturing rarest information in such modules has fluctuating performance either scoring the best in Top K (capturing the first error fastest) or APFD (capturing all the errors fastest). In summary, on average, that our method \textit{RSC} achieves the best performance across all modules in different test suite with $1.58$ for Top K and $0.61$ for APFD.

\begin{table*}
\centering
\caption{Effectiveness of the test reduction and prioritization methods in detecting faults}
\label{tab:exp2_1}
\arrayrulecolor{black}
\scalebox{1}{
\begin{tabular}{llcc|lllll|lllll} 
\hline
\rowcolor[rgb]{0.804,0.804,0.804} \multicolumn{1}{c}{{\cellcolor[rgb]{0.804,0.804,0.804}}}                                      & \multicolumn{1}{c}{{\cellcolor[rgb]{0.804,0.804,0.804}}}                               & {\cellcolor[rgb]{0.804,0.804,0.804}}                                        & {\cellcolor[rgb]{0.804,0.804,0.804}}                                                   & \multicolumn{5}{c|}{\textbf{\textbf{Top K}}}                                                                                                                                                                                                   & \multicolumn{5}{c}{\textbf{APFD}}                                          \\ 
\hhline{>{\arrayrulecolor[rgb]{0.804,0.804,0.804}}---->{\arrayrulecolor{black}}----------}
\rowcolor[rgb]{0.804,0.804,0.804} \multicolumn{1}{c}{\multirow{-2}{*}{{\cellcolor[rgb]{0.804,0.804,0.804}}\textbf{Test suite}}} & \multicolumn{1}{c}{\multirow{-2}{*}{{\cellcolor[rgb]{0.804,0.804,0.804}}\textbf{ADS Module}}} & \multirow{-2}{*}{{\cellcolor[rgb]{0.804,0.804,0.804}}\textbf{Total faults}} & \multirow{-2}{*}{{\cellcolor[rgb]{0.804,0.804,0.804}}\textbf{\textbf{Covered faults}}} & \textbf{CH} & \textbf{RD} & \textbf{CC}  & \textbf{SC}                                                                                           & \begin{tabular}[c]{@{}>{\cellcolor[rgb]{0.804,0.804,0.804}}l@{}}\textbf{RSC}\\\end{tabular} & \textbf{CH}   & \textbf{RD} & \textbf{CC} & \textbf{SC}   & \textbf{RSC}   \\ 
\hline
Cubetown                                                                                                                        & \multirow{4}{*}{Traffic light}                                                         & 13                                                                           & 13 (100\%)                                                                              & 6.0         & 4.26        & 5.5          & 5.0                                                                                                   & \textbf{2.0}                                                                                & 0.43          & 0.50        & 0.39        & 0.50          & \textbf{0.67}  \\
Gomentum                                                                                                                        &                                                                                        & 19                                                                          & 19 (100\%)                                                                             & 2.0         & 3.57        & 6.14         & \textbf{1.0}                                                                                          & \textbf{1.0}                                                                                & 0.50          & 0.50        & 0.25        & \textbf{0.82} & 0.76           \\
Shalun                                                                                                                          &                                                                                        & 78                                                                          & 72 (92.3\%)                                                                            & 1.83        & 1.46        & 2.17         & 2.0                                                                                                   & \textbf{1.33}                                                                               & \textbf{0.56} & 0.50        & 0.51        & \textbf{0.56} & 0.54           \\
\textbf{Total }                                                                                                                 &                                                                                        & \textbf{109}                                                                & \textbf{104 (95.4\%) }                                                                  & 3.28        & 3.10        & 4.60         & 2.67                                                                                                  & \textbf{1.44}                                                                               & 0.50          & 0.50        & 0.38        & 0.63          & \textbf{0.66}  \\ 
\hline
\rowcolor[rgb]{0.898,0.898,0.898} Cubetown                                                                                      & {\cellcolor[rgb]{0.898,0.898,0.898}}                                                   & 49                                                                          & 49 (100\%)                                                                             & 1.89        & 2.47        & \textbf{2.44}         & 3.0                                                                                                   & 3.0                                                                                         & 0.56          & 0.50        & 0.53        & \textbf{0.59} & \textbf{0.59}  \\
\rowcolor[rgb]{0.898,0.898,0.898} Gomentum                                                                                      & {\cellcolor[rgb]{0.898,0.898,0.898}}                                                   & 28                                                                          & 28 (100\%)                                                                             & 3.67        & 1.74        & 2.22         & \textbf{1.0}                                                                                          & \textbf{1.0}                                                                                & \textbf{0.51} & 0.50        & 0.34        & 0.44          & 0.47           \\
\rowcolor[rgb]{0.898,0.898,0.898} Shalun                                                                                        & {\cellcolor[rgb]{0.898,0.898,0.898}}                                                   & 27                                                                          & 27 (100\%)                                                                             & 2.0         & 1.17        & 1.22         & \textbf{1.0}                                                                                          & \textbf{1.0}                                                                                & 0.48          & 0.50        & 0.50        & \textbf{0.53} & \textbf{0.53}  \\
\rowcolor[rgb]{0.898,0.898,0.898} \textbf{ Total }                                                                              & \multirow{-4}{*}{{\cellcolor[rgb]{0.898,0.898,0.898}}Obstacle}                         & \textbf{94}                                                                 & \textbf{94 (100\%) }                                                                   & 2.52        & 1.79        & 1.96         & \begin{tabular}[c]{@{}>{\cellcolor[rgb]{0.898,0.898,0.898}}l@{}}\textbf{\textbf{1.67}}\\\end{tabular} & \textbf{1.67}                                                                               & 0.52          & 0.50        & 0.46        & 0.52          & \textbf{0.53}  \\ 
\hline
~Cubetown                                                                                                                       & \multirow{4}{*}{Planning~}                                                             & 42                                                                          & 42 (100\%)                                                                             & 3.0         & 1.38        & \textbf{1.0} & \textbf{1.0}                                                                                          & \textbf{1.0}                                                                                & 0.37          & 0.50        & 0.58        & \textbf{0.59} & \textbf{0.59}  \\
Gomentum                                                                                                                        &                                                                                        & 10                                                                          & 10 (100\%)                                                                             & 7.0         & 4.48        & 7.14         & 7.0                                                                                                   & \textbf{2.0}                                                                                & 0.19          & 0.50        & 0.17        & 0.19          & \textbf{0.81}  \\
Shalun                                                                                                                          &                                                                                        & 64                                                                          & 64 (100\%)                                                                             & 3.0         & 1.60        & \textbf{1.0} & \textbf{1.0}                                                                                          & 1.44                                                                                        & 0.49          & 0.50        & 0.56        & 0.53          & \textbf{0.58}  \\
\textbf{ Total }                                                                                                                &                                                                                        & \textbf{116}                                                                & \textbf{116 (100\%) }                                                                  & 4.33        & 2.49        & 3.05         & 3.0                                                                                                   & \textbf{1.48}                                                                               & 0.35          & 0.50        & 0.44        & 0.44          & \textbf{0.66}  \\ 
\hline
\rowcolor[rgb]{0.898,0.898,0.898} Cubetown                                                                                      & {\cellcolor[rgb]{0.898,0.898,0.898}}                                                   & 34                                                                          & 34 (100\%)                                                                             & 4.44        & 3.27        & 4.44         & \textbf{2.22}                                                                                         & \textbf{2.22}                                                                               & 0.39          & 0.50        & 0.48        & 0.61          & \textbf{0.62}  \\
\rowcolor[rgb]{0.898,0.898,0.898} Gomentum                                                                                      & {\cellcolor[rgb]{0.898,0.898,0.898}}                                                   & 61                                                                          & 61 (100\%)                                                                             & 1.22        & 1.13        & 1.11         & \textbf{1.0}                                                                                          & \textbf{1.0}                                                                                & 0.50          & 0.50        & 0.50        & 0.50          & \textbf{0.51}  \\
\rowcolor[rgb]{0.898,0.898,0.898} Shalun                                                                                        & {\cellcolor[rgb]{0.898,0.898,0.898}}                                                   & 19                                                                          & 19 (100\%)                                                                             & 4.63        & 5.82        & 6.50         & 3.75                                                                                                  & \textbf{2.0}                                                                                & 0.55          & 0.50        & 0.45        & 0.61          & \textbf{0.64}  \\
\rowcolor[rgb]{0.898,0.898,0.898} \textbf{ Total }                                                                              & \multirow{-4}{*}{{\cellcolor[rgb]{0.898,0.898,0.898}}Prediction}                       & \textbf{114}                                                                & \textbf{114 (100.0\%) }                                                                & 3.43        & 3.41        & 4.02         & 2.32                                                                                                  & \textbf{1.74}                                                                               & 0.48          & 0.50        & 0.48        & 0.57          & \textbf{0.59}  \\ 
\hline
\textbf{Total}                                                                                                                  & \multicolumn{1}{c}{\textbf{Total}}                                                     & \textbf{433}                                                                & \textbf{428(98.8\%)}                                                                   & 3.39        & 2.70        & 3.41         & 2.42                                                                                                  & \textbf{1.58}                                                                               & 0.46          & 0.50        & 0.44        & 0.54          & \textbf{0.61}  \\
\hline
\end{tabular}}
\end{table*}

\section{Related Work}
\label{sec:related_work}
\subsection{Test Reduction and Prioritization}
Test reduction and prioritization are two approaches to reduce the cost of regression testing. Test reduction, also called test minimization~\cite{rothermel1998empirical}, seeks to reduce the size of a test suite by removing redundant test cases. Test prioritization~\cite{rothermel1999test} aims to maximize some desired properties such as the rate of fault detection by sorting test cases. The comprehensive overview of regression testing techniques can be checked in the survey~\cite{yoo2012regression}.

The typical test reduction techniques contain heuristic methods~\cite{harrold1993methodology, chen1998new, chen1970heuristics, chen1998simulation} and genetic algorithm-based approaches~\cite{mansour1999simulated, ma2005test}. Different heuristics and search algorithms were applied to select a minimal set of test cases that achieve the same test requirements (e.g., branch coverage and code coverage) as the original test suite. For test prioritization, many coverage-based prioritization methods~\cite{rothermel2001prioritizing, elbaum2000prioritizing, elbaum2001incorporating} were proposed to maximize structural coverage or code coverage early. The idea behind these methods is that the early maximization of structure coverage will improve the chance to detect faults early~\cite{yoo2012regression}. These existing test reduction and prioritization methods target on traditional programs. However, they cannot be directly applied on ADSs with time series inputs. Inspired by traditional coverage metrics, in this paper we propose rarity and semantic coverage-based test prioritization method.  

There are a few works related to test reduction and prioritization on ADSs. In~\cite{lu2021search}, Lu et al.~conducted experiments to evaluate the performances of five search algorithms for prioritizing driving scenarios and defined objective functions (e.g., the probability of collision) for optimization. However, it does not account for the coverage of different modules in an ADS, which limits its prioritization capability.  In our work, the test reduction process is automated based on frame vectorization calculated by the data in driving recordings. Furthermore, we evaluate our method on four ADS modules. In~\cite{birchler2021automated}, Birchler et al.~proposed a prioritization method to sort driving scenarios of lane-keeping systems. In their work, testing suites are maps with different shapes, turns, and other properties. They extracted features of maps and sorted them using genetic algorithms. In our work, testing suites are different driving recordings and the testing modules are perception, prediction, and planning modules. We split driving recordings to recording segments for further test reduction and prioritization. 

\subsection{Autonomous Driving Testing}
The research of testing on ADSs mainly focused on the generation of rare or error-prone driving scenarios.  In~\cite{tian2018deeptest}, Tian et al.~proposed to apply different affine transformations to generate new testing images to evaluate the steering angle prediction of CNN-based driving models. In~\cite{zhang2018deeproad}, Zhang et al.~ proposed to use Generative adversarial networks (GANs) to generate high-quality testing images such as driving scenes on rainy and snowy days. In~\cite{deng2021bmt}, Deng et al.~proposed to generate driving scenarios based on traffic rules to test driving models for speed prediction. In~\cite{zhou2020deepbillboard}, Zhou et al.~ proposed to use adversarial attack methods~\cite{deng2021deep,deng2020analysis} to generate adversarial billboards in driving scenarios. In~\cite{gambi2019generating},  Gambi et al.~generated crash driving scenarios based on police reports. 

Several search-based methods were proposed to find and generate driving scenarios leading to crash or deviation of ego-vehicles~\cite{abdessalem2018testing, ben2016testing, gambi2019automatically, calo2020generating, li2020av, ebadi2021efficient,abdessalem2018testing2, gambi2019asfault}.
While most work targeted on E2E driving models or Advanced driver-assistance systems (ADAS), some recent works started investigating on industry-level ADSs. In~\cite{garcia2020comprehensive}, Garcia et al.~systematically analyzed bugs in Apollo and Autoware based on their Github repository commits.  In~\cite{li2020av}, Li et al.~proposed AV-FUZZER to search safety violations of Apollo in the urban driving environment using genetic algorithms~\cite{kumar2010genetic}. In~\cite{ebadi2021efficient}, Ebadi et al.~searched testing scenarios for obstacle detection of Apollo. Our work focus on test reduction and prioritization based on driving recordings.

\subsection{Driving Scenario Identification}
Driving scenario identification is the task of identifying driving scenarios and converting driving scenarios to vectors. After this, a few downstream works such as scenario clustering and anomaly detection can be implemented. Currently, only a few works in the software engineering community researched in this direction. In~\cite{stocco2020misbehaviour}, Andrea et al.~proposed a method to detect potential misbehaviors of an ADS. The main idea is to train a reconstructor to reconstruct the current driving scenario (single image or a sequence of images). If the reconstruction error exceeds a threshold, the driving scenario is most likely an anomalous scenario, and the autonomous vehicle may misbehave in the scenario.  In this work, we identify driving scenarios based on the module outputs in the driving recording and convert each frame to a feature vector. We then use feature vectors for testing reduction and prioritization.


Beyond SE community, some existing work targeted driving scenario vectorization. In~\cite{harder2021scenario2vector}, Harder et al.~proposed \textit{scenario2vector} to characterize a driving image from three aspects \textit{actors, actions}, and \textit{scene}. Then based on the text description of the driving image, key elements belonging to the above three categories are extracted and converted to vectors or matrices. In our work, we also describe key elements of driving scenarios from these three aspects. However, we use them to define a driving scene schema to describe corresponding semantic information occurred in a driving scene. On the other hand, we do not use them together to characterize a driving scenario. For different testing modules, We extract data from its publish channel messages to encode semantic information described in the schema.    

In~\cite{zhao2021large}, Zhao et al.~proposed using Convolutional Neural Network (CNN) and Gated recurrent unit (GRU) network~\cite{chung2014empirical} to learn feature representations for driving videos. 
In~\cite{balasubramanian2021traffic}, Balasubramanian et al.~proposed to use self-supervised learning~\cite{jing2020self} to learn better feature representations for driving recordings.  In our work, we treat a driving recording as a combination of multiple driving scenarios. We create and calculate frame-level features for driving recording segmentation. 

\section{Threats to Validity}
\label{sec:threat}
We discuss threats to validity from three aspects external validity, construct validity, and internal validity proposed in~\cite{Experimentation}.

\textbf{External Validity:}
The external validity is generalizability of the proposed method. In this work, we conduct experiments on an industry-level ADS Apollo to evaluate the effectiveness of the proposed test reduction and prioritization methods. 
These methods are also available for other multi-module ADSs that apply publish-subscribe communication mechanism among modules and save module outputs in driving recordings (e.g., Autoware~\cite{kato2018autoware} that adopts ROS~\cite{afzal2020study}). 

\textbf{Internal Validity:}
The internal validity is that in ADS modules there are many non-deterministic algorithms, which makes module outputs have slight differences when the module runs on the same input multiple times~\cite{koopman2016challenges}. This problem may introduce uncertain faults when we compare the outputs of the old and new versions of SUTs. To solve the problem, We use a sliding window-based smoothing algorithm to remove noises. Furthermore, we set the threshold to evaluate inconsistent outputs as 10\% to ensure the detected inconsistent output is caused by the new version of SUT. In this work, we inject mutants into Apollo source code to create regression benchmarks. The effectiveness of mutation testing are suitable for software testing experimentation~\cite{andrews2005mutation, just2014mutants}. In this way, we explore common faults in different ADS modules and ensure every time we only change one file in an ADS module. We do not use commits in Apollo Github commit history as regression benchmarks because most commits are relevant to the update of multiple modules.  

\textbf{Construct Validity:}
The construct validity is that we conducted experiments on Apollo 5.0, not Apollo 6.0 that is used in other papers~\cite{li2020av}. The reason is that now for Apollo 6.0, the camera-based perception module does not work and LiDAR-based perception is unstable in SVL
\footnote{\url{https://bit.ly/3bVX2LR}}.
In this work, we evaluated our methods on three maps. We tested two other maps (i.e., `Borregas Avenue' and `San Francisco') but we found that Apollo is not able to control the ego-vehicle stably. This issue has been reported by Apollo users
\footnote{\url{https://bit.ly/3zUGmMP}}\footnote{\url{https://bit.ly/3di6slc}}. Therefore, we excluded these two maps from the experiment. We will keep tracking these issues and try to include more maps for experiments in the future work. For mutant rejection, we did not directly mutate DL models in Apollo because the models are encapsulated in binary files. Alternatively, we mutated C++ files that are related to DL model configuration and model input/output to simulate bugs in DL models. For example, by mutating the output of the obstacle detection model, we can simulate incorrect model predictions.


\section{Conclusion}
\label{sec:conclusion}
This paper proposes {\tool} to reduce the cost of regression testing in industry-scale multi-module ADSs. Since such ADSs are largely built upon publish-subscribe mechanism, we define a driving scene schema and for different modules under test, we extract frame data from its publish channels and encode semantic information described in the schema as frame feature vectors. Based on such feature vectorization, we propose a test reduction algorithm to split original driving recordings to recording segments, each of which reflects a unique driving scenario. We also propose semantic coverage based and rarity based test prioritization to order the reduced recording segments as test inputs for ADS modules. In our work, testing suites are different driving recordings and the test modules are perception, prediction, and planning modules in ADSs. In our evaluation using an industry-level multi-module ADS and three road maps in diverse regression testing settings, we prove {\tool} is able to significantly reduce the length of original driving recordings ($34\%$ to $77\%$), has promising fault coverage of $98.8\%$, and achieves $1.58$ for Top K and $0.61$ for APFD beating the state-of-the-art baselines. This work is a pioneering work to utilize frame-level features for driving recording segmentation. With manually defined driving scene feature schema, we have achieved promising results in both test reduction and prioritization. As future directions for SE community, based on this open-sourced work, automated feature extraction from driving recording can be explored along with more intelligent smoothing algorithms.

\section{Acknowledgements}
This work is in part supported by an Australian Research Council (ARC) Discovery Project (DP210102447), an ARC Linkage Project (LP190100676), and a DATA61 project (Data61 CRP C020996).

\bibliographystyle{ACM-Reference-Format}
\balance
\bibliography{reference.bib}

\end{document}